\documentclass{article}
\usepackage{amsfonts,amssymb,xspace}
\newlength{\dinwidth}
\newlength{\dinmargin}
\makeatletter
\newif\if@fewtab\@fewtabtrue
\@addtoreset{equation}{section}
\def\cases#1{\left\{\,\vcenter{\normalbaselines\m@th
    \ialign{$\displaystyle{##}\hfil$&\quad##\hfil\crcr#1\crcr}}\right.}
\def\casetable#1{\left\{\,\vcenter{\normalbaselines\m@th
    \ialign{$\hfil\displaystyle{##}$&\quad##\hfil\crcr#1\crcr}}\right.}
\def\sideset#1#2#3{%
  \setbox\z@\hbox{$\displaystyle{\vphantom{#3}}#1{#3}\m@th$}%
  \setbox\tw@\hbox{$\displaystyle{#3}#2\m@th$}%
  \hskip\wd\z@\hskip-\wd\tw@\mathop{\hskip\wd\tw@\hskip-\wd\z@
  {\vphantom{#3}}#1\!{#3}#2}}
\makeatother
\newcommand{\Cn}{\mathbb{C}}

\newcommand{\Zn}{\mathbb{Z}}
\newcommand{\Nn}{\mathbb{N}}

\newtheorem{Thm}{Theorem}
\newtheorem{Cor}{Corollary}

\newcommand{\lb}[1]{\label{#1}}
\newcommand{\Eq}[1]{(\ref{#1})}
\newcommand{\ct}[1]{\cite{#1}}

\renewcommand{\[}{\begin{eqnarray}}
\newcommand{\nn}{\nonumber}
\newcommand{\non}{\nonumber \\ }
\renewcommand{\]}{\end{eqnarray}}
\newcommand{\hh}[1]{\hspace*{#1}}

\newcommand{\een}{\end{enumerate}}
\newcommand{\ben}{\begin{enumerate}}

\newcommand{\noi}{\noindent}
\newcommand{\ga}{\alpha}

\newcommand{\gd}{\delta}
\newcommand{\gD}{\Delta}
\newcommand{\gep}{\epsilon}

\newcommand{\gl}{\lambda}
\newcommand{\gL}{\Lambda}

\newcommand{\go}{\omega}
\newcommand{\gO}{\Omega}
\newcommand{\gp}{\psi}

\newcommand{\gt}{\theta}
\newcommand{\gx}{\xi}

\newcommand{\gz}{\zeta}
\newcommand{\cl}{\ensuremath{\ell}\xspace}

\newcommand{\cF}{\mathcal{F}}

\newcommand{\cP}[1][]{\mathcal{P}^{(#1)}}
\newcommand{\cPL}{\ensuremath{\mathcal{P}(\vgL)}}

\newcommand{\fg}{\ensuremath{\mathfrak{g}}\xspace}
\newcommand{\fh}{\mathfrak{h}}
\newcommand{\fn}{\mathfrak{n}}

\newcommand{\fT}{\mathfrak{T}}

\newcommand{\fW}{\mathfrak{W}}

\newcommand{\rP}{\mathrm{P}}

\newcommand{\sP}{\mathcal{P}}
\newcommand{\sX}{\mathcal{X}}

\newcommand{\va}{\mathbf{a}}
\newcommand{\vga}{\mbox{\boldmath$\ga$}}

\newcommand{\vgd}{\mbox{\boldmath$\gd$}\xspace}

\newcommand{\vk}{\mathbf{k}}

\newcommand{\vgl}{\mbox{\boldmath$\gl$}\xspace}
\newcommand{\vgL}{\mbox{\boldmath$\gL$}\xspace}
\newcommand{\vo}{\mathbf{0}}
\newcommand{\vp}{\mathbf{p}}
\newcommand{\vP}{\mathbf{P}}
\newcommand{\vq}{\mathbf{q}}
\newcommand{\vQ}{\mathbf{Q}}
\newcommand{\vX}{\mathbf{X}}
\newcommand{\vr}{\mathbf{r}}
\newcommand{\vs}{\mathbf{s}}

\newcommand{\vgt}{\mbox{\boldmath$\gt$}}

\newcommand{\vv}{\mathbf{v}}

\newcommand{\vgx}{\mbox{\boldmath$\gx$}}

\newfont\sgb{cmmib7} 

\newcommand{\svt}{{\hbox{\sgb \symbol{"12}}}}
\newcommand{\svd}{{\hbox{\sgb \symbol{"0E}}}}

\newcommand{\svl}{{\hbox{\sgb \symbol{"15}}}}
\newcommand{\svL}{{\hbox{\sgb \symbol{"03}}}}
\newcommand{\Oint}{\oint\limits}

\newcommand{\Res}[2][]{\Oint_{#1}\!\frac{d#2}{2\pi i}\,}
\newcommand{\End}{\mathrm{End}}

\renewcommand{\span}{\mathrm{span}}

\newcommand{\mult}{\mathrm{mult}}

\newcommand{\re}{\mathrm{e}}
\newcommand{\frc}[2]{{\textstyle \frac{#1}{#2}}}

\newcommand{\dz}{\frac{d}{dz}}

\renewcommand{\.}{\cdot}
\newcommand{\X}{\!\cdot\!}
\newcommand{\XO}{\otimes}

\newcommand{\ket}[1]{\ensuremath{|#1\rangle}}

\newcommand{\ord}[1]{\mbox{\large\bf:} #1 \mbox{\large\bf:}}
\newcommand{\xord}[1]{{}_\times^\times #1 {}_\times^\times}
\newcommand{\circord}[1]{{}_\circ^\circ #1 {}_\circ^\circ}

\newcommand{\hv}{\ensuremath{h^\vee}\xspace}
\newcommand{\bfg}{\ensuremath{\bar{\fg}}\xspace}
\newcommand{\bfgh}{\ensuremath{\hat{\fg}}\xspace}
\newcommand{\bfh}{\ensuremath{\bar{\fh}}\xspace}

\newcommand{\bD}{\ensuremath{\bar{\Delta}}\xspace}

\newcommand{\vkl}{\ensuremath{\vk_\cl}\xspace}

\newcommand{\AI}[2]{\ensuremath{A^{#1}_{#2}}}
\newcommand{\Ai}[1]{\ensuremath{A^i_{#1}}}
\newcommand{\Aj}[1]{\ensuremath{A^j_{#1}}}

\newcommand{\AL}[1]{\ensuremath{A^-_{#1}}}
\newcommand{\AP}[1]{\ensuremath{A^+_{#1}}}
\newcommand{\Ar}[1]{\ensuremath{A^{\vr}_{#1}}}
\newcommand{\Er}[1]{\ensuremath{E^{\vr}_{#1}}}
\newcommand{\Es}[1]{\ensuremath{E^{\vs}_{#1}}}
\newcommand{\Erb}[1]{\ensuremath{E^{\vr+\vs}_{#1}}}
\newcommand{\Ema}[1]{\ensuremath{E^{-\vr}_{#1}}}
\newcommand{\Ehr}[1]{\ensuremath{\hat{E}^{\vr}_{#1}}}
\newcommand{\Ehs}[1]{\ensuremath{\hat{E}^{\vs}_{#1}}}
\newcommand{\Ehrs}[1]{\ensuremath{\hat{E}^{\vr+\vs}_{#1}}}
\newcommand{\Ehmr}[1]{\ensuremath{\hat{E}^{-\vr}_{#1}}}
\newcommand{\XI}[1]{\ensuremath{\sX^i_{#1}}}
\newcommand{\XL}[2]{\ensuremath{\sX^{#1}_{#2}}}
\newcommand{\vcX}{\ensuremath{\sX}}
\newcommand{\PI}[1]{\ensuremath{\sP^i_{#1}}}
\newcommand{\PL}[2]{\ensuremath{\sP^{#1}_{#2}}}
\newcommand{\vcP}{\ensuremath{\sP}}

\newcommand{\sL}[2][\cl]{\ensuremath{\mathcal{L}^{\scriptscriptstyle%
                                                  [#1]}_{#2}}}

\newcommand{\piL}{\pi_{\svL}}
\newcommand{\tr}[2][\cl]{{\sideset{^{\scriptscriptstyle[#1]}}{_{#2}}{t}}}
\newcommand{\sT}[1][\cl]{{\sideset{^{\scriptscriptstyle[#1]}}%
                                             {}{\mathcal{T}}}}
\newcommand{\hsT}[1][\cl]{{\sideset{^{\scriptscriptstyle[#1]}}%
                                             {}{\widehat{\mathcal{T}}}}}
\newcommand{\gol}[1][\cl]{{\sideset{^{\scriptscriptstyle[#1]}}{}{\go}}}

\begin{document}
\arraycolsep3pt
\pagestyle{myheadings}
\markboth{R.~W.~Gebert and H.~Nicolai: Affine string vertex operators at
arbitrary level}{R.~W.~Gebert and H.~Nicolai: Affine string vertex
operators at arbitrary level}
\thispagestyle{empty}
\begin{flushright} hep-th/9608014
               \\  DESY 96-166  \end{flushright}
\vspace*{2cm}
\begin{center}
 {\LARGE \sc An Affine String Vertex Operator\\[1ex]
            Construction at Arbitrary Level%
            \footnote[1]{To appear in \emph{J. Math. Phys.}}
  }\\
 \vspace*{1cm}
 {\sl
     R.~W.~Gebert\footnotemark[2]$^,$%
      \footnote[3]{Supported by Deutsche Forschungsgemeinschaft
                   under Contract No.\ \emph{DFG Ge 963/1-1}}\qquad
     H. Nicolai\footnotemark[4]}\\
 \vspace*{6mm}
     \footnotemark[2]
     Institute for Advanced Study, School of Natural Sciences\\
     Olden Lane, Princeton, NJ 08540, U.S.A.\\
 \vspace*{3mm}
     \footnotemark[4]
     Max-Planck-Institut f\"ur Gravitationsphysik \\
     Albert-Einstein-Institut \\
     Schlaatzweg 1, D-14473 Potsdam, Germany\\
 \vspace*{6mm}
 \centerline{29 April 1997}
 \vspace*{1cm}
 \begin{minipage}{11cm}\footnotesize
   {\bf Abstract:} An affine vertex operator construction at arbitrary
   level is presented which is based on a completely compactified
   chiral bosonic string whose momentum lattice is taken to be the
   (Minkowskian) affine weight lattice. This construction is
   manifestly physical in the sense of string theory, i.e., the vertex
   operators are functions of DDF ``oscillators'' and the Lorentz
   generators, both of which commute with the Virasoro constraints. We
   therefore obtain explicit representations of affine highest weight
   modules in terms of physical (DDF) string states. This opens new
   perspectives on the representation theory of affine Kac--Moody
   algebras, especially in view of the simultaneous treatment of
   infinitely many affine highest weight representations of arbitrary
   level within a single state space as required for the study of
   hyperbolic Kac--Moody algebras.  A novel interpretation of the
   affine Weyl group as the ``dimensional null reduction'' of the
   corresponding hyperbolic Weyl group is given, which follows upon
   re-expression of the affine Weyl translations as Lorentz boosts.
 \end{minipage}
\end{center}
 \vspace*{1cm}
 \centerline{PACS: 11.25.Hf; 02.20.Sv; 02.20.Tw}
\setcounter{footnote}{0}
\newpage

\section{Introduction} \lb{sec:Int}
In this paper we propose a generalization of the Frenkel--Kac--Segal
(FKS) vertex operator realization of nontwisted affine Lie algebras at
level one \ct{FreKac80,Sega81} to arbitrary level. This construction
was originally based on the spatial compactification of a bosonic
string whose momentum lattice is taken to be the (Euclidean) root
lattice of a finite-dimensional simple Lie algebra of $ADE$ type. The
Laurent coefficients (modes) of the tachyon vertex operators together
with the string oscillators then constitute a basis of the affine
algebra. This basis is not physical in the sense of string theory
since, except for the zero mode, these operators do not commute with
the Virasoro constraints. However, there is also a ``covariant''
version of the FKS construction \ct{Fren85,GodOli85}, where the
momentum lattice of the string is enlarged by a two-dimensional
Minkowski lattice; then the zero mode operators are indeed physical in
the sense of string theory and already by themselves form a basis of
the affine algebra. Apparently it has not been generally appreciated
so far that, apart from being manifestly physical, this construction
is applicable to affine Lie algebras at arbitrary level and thus more
general than the FKS construction. The characteristic feature of our
model is that the momentum lattice of the string is taken to be the
(Minkowskian) affine weight lattice. This model was recently exploited
in \ct{GeKoNi96} to construct an explicit representation of the affine
Sugawara generators in terms of (transversal) DDF operators at
arbitrary level.

A main new result of this paper is a string vertex operator
realization of the affine Cartan--Weyl basis (in particular the step
operators) at arbitrary level in terms of physical (DDF) operators
rather than ordinary string oscillators as in \ct{Fren85,GodOli85}.
Consequently, we can evaluate the action of these operators on any
given physical state directly in terms of the DDF basis.  This
construction leads us to consider a new type of (level-dependent)
physical field $\sX^\mu(z)$, similar to the old Fubini-Veneziano
field, $X^\mu(z)$, but where the ordinary string oscillators are
replaced by level-\cl transversal DDF oscillators. Apart from the
center of mass coordinate, the fields $\sX^\mu$ were already
introduced in \ct{GeKoNi96}. Whereas the correct definition of the
center of mass mode of $\sX^\mu$ did not matter in \ct{GeKoNi96}, it
is absolutely essential here; somewhat surprisingly, this definition
turns out to involve the Lorentz generators $M_{\mu \nu}$, which are
physical operators, rather than the operators $q^\mu$, which are not
physical unlike the center of mass momenta $p^\mu$. The construction
also requires a corresponding new type of ``tachyon vertex operator"
with $X^\mu$ replaced by $\sX^\mu$; more general operators of this
type presumably will be needed at a later stage.

The proper definition of the center of mass mode, and the
corresponding replacement of a translation generator by a Lorentz
transformation leads us to our second main result, namely a new
interpretation of the affine Weyl group as a ``dimensional null
reduction'' of the hyperbolic Weyl group. More specifically, this
result hinges on re-expressing the so-called (affine) Weyl
translations as Lorentz boosts. Consequently, these elements of the
affine Weyl group should really be called ``Weyl boosts''. In this way
it becomes obvious that the embeddings of the finite, affine and
hyperbolic Weyl groups of the finite, affine and hyperbolic Kac--Moody
algebras $\bfg\subset\fg\subset\bfgh$, respectively (with the finite
algebra $\bfg$ of rank $d-2$), are just the discrete analogs of the
corresponding sequence of embeddings of the continuous groups
$\mathrm{SO}(d-2)\subset \mathrm{ISO}(d-2)\subset \mathrm{SO}(d-1,1)$
into one another (see diagram \Eq{diagram}). Here $\mathrm{ISO}(d-2)$
is defined to be the subgroup of $\mathrm{SO}(d-1,1)$ leaving
invariant a given light-like vector, which in our case is just the
affine null root $\vgd$. We find it remarkable that this new
description of the affine Weyl group is really forced upon us by the
DDF approach, and this suggests that it is the truly natural
interpretation of the known result that the affine Weyl group is a
semidirect product of the finite Weyl group and the affine Weyl
translations. While the affine Weyl transformations leave the level of
a given representation fixed, one can in principle also consider
level-changing boosts. As we expect such transformations to provide
important new insights into the structure of hyperbolic Kac--Moody
algebras and their Weyl groups we briefly discuss these
generalizations in the last section.

We believe that the results presented in this paper open new and
promising perspectives for the theory of irreducible representations
of affine Lie algebras, especially with regard to the problem of
understanding hyperbolic Kac--Moody algebras (actually our main goal),
where one must simultaneously deal with infinitely many inequivalent
representations of arbitrary level. As shown in the present paper,
this aim can be achieved by embedding all the affine representation
spaces into a single Fock space of physical states.  Among the
fascinating open problems for further study let us especially mention
the idea of extending the present construction to ``level-changing
vertex operators" (actually, this will be a generic feature when we go
over to the hyperbolic extension of the affine Lie algebra), possibly
also of more general type than the tachyon-type vertex operators
utilized here.  Whereas the affine generators themselves involve only
transversal DDF operators and thus contribute only transversal
excitations to states within a single irreducible affine
representation, the longitudinal DDF operators by construction map
affine vacuum vectors into each other and will accordingly act as
representation-changing (and in general even level-changing)
operators. The latter have so far played no role in the representation
theory of affine algebras, and are unnecessary as long as one deals
only with one representation at a time.  However, it is clear
that a proper understanding of the longitudinal DDF operators is
one of the keys to unraveling the mysteries of indefinite and
hyperbolic Kac--Moody algebras.

\section{Affine weights and DDF operators} \lb{sec:Set}
We consider a nontwisted affine Lie algebra
$\fg=\fn_-\oplus\fh\oplus\fn_+$ (for general references on this
subject see e.g.\ \ct{Kac90,MooPia95}) with underlying simple
finite-dimensional Lie algebra \bfg of type $ADE$ and with rank $d-2$
($d>2$). The affine (resp.\ finite) root lattice is denoted by $Q$
(resp.\ $\bar{Q}$). The space of dominant integral affine weights is
given by
\[ \rP_+:=\{\vgL\in\fh^*|\vgL\X\vr_I\in\Zn_+, 0\le I\le d-2\}
         =\sum_{I=0}^{d-2}\Zn_+\vgL_I+\Cn\vgd, \nn \]
where $\vgd$ is the affine null root, $\vr_I$ are the affine simple
roots, and $\vgL_I$ ($0\le I\le d-2$) are the fundamental affine
weights defined by $\vgL_I\X\vr_J=\gd_{IJ}$ and $\vgL_I\X\vgL_0=0$ for
$0\le I,J\le d-2$. For any dominant weight \vgL the level~$\cl\in\Zn$
is defined by\footnote{In comparison with \ct{GeKoNi96} we have
  switched signs so as to be in unison with the standard conventions
  \ct{Kac90,MooPia95}.}
\[ \cl:=\vgL\X\vgd. \]
By $L(\vgL)$ we denote the corresponding integrable irreducible
highest weight module over \fg. It is clear that $L(\vgL)\cong
L(\vgL+z\vgd)$ for all $z\in\Cn$. By putting
\[ \vgL':=\vgL+\frac1\cl\left(1-\frc12\vgL^2\right)\vgd \nn \]
for any $\vgL\in\rP_+$ of nonzero level we thus obtain a ``tachyonic''
(i.e.\ $\vgL'^2=2$) dominant integral affine weight which gives rise
to a highest weight module $L(\vgL')$ isomorphic to $L(\vgL)$. Without
loss of generality we shall assume from now on that $\vgL$ is some
tachyonic dominant weight of positive level~\cl.

Now let $\vgl$ be any weight in $\gO(\vgL)$, the set of weights for
$L(\vgL)$. It ensues that (i) $\vgl^2\in2\Zn$ and (ii)
$\vgl^2\le\vgL^2=2$. To see this we note that $\vgl=\vgL-\vr$ for some
$\vr\in Q_+$. Then (i) follows from $\vgl^2= 2(1-\vgL\X\vr)+ \vr^2$ by
the fact that $\vgL$ is an integral weight and that $Q$ is an even
lattice by assumption. To prove (ii), we use that $\vgl$ is
Weyl-equivalent to a unique $\vgl'\in\rP_+\cap\gO(\vgL)$ with
decomposition $\vgl'=\vgL-\vr'$ for some $\vr'\in Q_+$ ; hence
$\vgl^2= {\vgl'}^2 =\vgL^2- \vgL\X\vr'- \vgl'\X\vr'\le \vgL^2= 2$
because both $\vgL$ and $\vgl'$ are dominant (cf.\ \cite[Prop.\
11.4.a)]{Kac90}). 

These observations are crucial for the DDF construction to be 
described below: for any level-\cl weight $\vgl\in\gO(\vgL)$ 
we define its DDF decomposition \ct{GebNic95} by
\[ \vgl = \va - n \vkl, \lb{DDF-dec} \]
where
\[ \vkl := \frac1\cl\vgd, \lb{k-def} \]
and the vector $\va$ is uniquely determined by demanding $\va^2=2$. 
Thus $n=1-\frc12\vgl^2$, and by the above result, $n$ is
always a non-negative integer as required by the DDF construction.
We will refer to $\va$ as the ``tachyonic level-\cl vector'' and to the 
corresponding state $\ket\va$ as the ``tachyonic level-\cl state'' 
associated to \vgl. Note that, for $\cl >1$, the tachyonic vector
$\va$ occurring in Eq.\ \Eq{DDF-dec} in general is \emph{not} a weight
for $L(\vgL)$ because of the fractional coefficient in front of $\vgd$.
Rather it will be used as an auxiliary vector in
the construction below.

A central feature of our approach is the realization of the
affine representation space $L(\vgL)$ as a (tiny) subspace of 
a much bigger space $\cal P$ of physical string states, 
itself a subspace of a Fock space $\cF$ which is the direct sum
of irreducible Heisenberg modules modules created by the usual
string oscillators from the ground states $\ket{\vgl}\equiv
\exp (i\vgl\X {\bf q}) \ket{\vo}$ for arbitrary 
affine weights $\vgl$. More precisely,
\[ \cF := \span \left\{ \ga^{\mu_1}_{-m_1}\cdots \ga^{\mu_M}_{-m_M}
   \ket{\vgl} \, \big| \, \vgl\in\fh^* , m_\mu >0 \right\} \lb{Fock} \]
where the string oscillators $\ga^\mu_m$ ($m\in\Zn$,
$0\le\mu\le d-1$) and the center of mass operators $q^\mu , p^\mu$ obey
the standard commutation relations
\[ [\ga^\mu_m,\ga^\nu_n]=m\eta^{\mu\nu}\gd_{m+n,0},\qquad
   [q^\mu,p^\nu]=i\eta^{\mu\nu},\qquad \ga^\mu_0\equiv p^\mu. \nn \]
with $p^\mu \ket{\vgl} = \gl^\mu \ket{\vgl}$. 
To isolate the physical states, we introduce the Virasoro operators
\[ L_m:=\frac12\sum_{n\in\Zn}\ord{\vga_n\X\vga_{m-n}}, \lb{Vir} \]
which satisfy a Virasoro algebra with central charge $c=d$ (the
normal-ordering $\ord\ldots$ w.r.t. the string oscillators
$\ga^\mu_m$ in \Eq{Vir} is defined in the usual manner). The space of
physical string states ${\cal P}\subset\cF$ is then defined as 
\[ {\cal P} := \left\{ \psi \in\cF \, | \, L_0 \psi =\psi, L_n\psi =0 \; 
    \forall n>0 \right\}       \lb{Phys} \]
As already indicated, we shall be interested in certain subspaces of 
$\cal P$; more specifically, the affine representation space $L(\vgL)$ 
associated with the highest weight $\vgL$ will be embedded 
into the space
\[ \cPL=\bigoplus_{\svl\in\gO(\svL)}\cP[\svl]\subset {\cal P}, \]
where
\[ \cP[\svl]:=\{\gp\in\cF\|L_0\gp=\gp,\ L_n\gp=0\ \forall n>0,
                \ p^\mu\gp=\gl^\mu\gp, 0\le\mu\le d-1 \} \lb{phys-def} \]
denotes the space of physical string states with momentum $\vgl$. 

An explicit realization of the physical states is afforded by
the so-called DDF operators \ct{DeDiFu72,Brow72}.
To write them down we need the DDF decomposition \Eq{DDF-dec} since
these operators will always act on some tachyonic state $\ket{\va}$
associated with a given weight $\vgl$ in the sense explained above.
Furthermore, we need a set of polarization vectors
$\vgx^i\equiv\vgx^i(\va)\equiv\vgx^i(\vgl)$ ($1\le i\le d-2$)
satisfying $\vgx^i\X\vgx^j=\gd_{ij}$ and $\vgx^i\X\vgd=\vgx^i\X\va=0$,
which constitute a basis for the complex vector space $\bfh^*$ dual to
the Cartan subalgebra $\bfh$ of \bfg. The DDF operators are defined by
\ct{DeDiFu72,Brow72}\footnote{To make the notation less cumbersome,
  and contrary to the notational conventions of \ct{GeKoNi96}, we here
  suppress the label \cl on the DDF operators in \Eq{t-DDF} and
  \Eq{l-DDF} because this dependence is already implied by their
  dependence on $\va$.}
\[ \Ai{m}(\va)
    &:=& \Res{z}\vgx^i(\va)\X\vP(z)
                \exp\left[im\vkl\X\vX(z)\right], \lb{t-DDF} \\
   \AL{m}(\va)
    &:=& \Res{z}\ord{\left[-\va\X\vP(z)
           +\frac{m}2\frac{d}{dz}\ln\vkl\X\vP(z)\right]
      \exp\left[im\vkl\X\vX(z)\right]} - \non
    & &  -\frac12 \sum_{n\in\Zn} \xord{\Ai{n}(\va)\Ai{m-n}(\va)}
         +2\gd_{m0}\vkl\X\vp,
   \lb{l-DDF} \\
   \AI+m(\va)
    &:=& \Res{z}\vkl\X\vP(z)\exp\left[im\vkl\X\vX(z)\right]
      = \gd_{m0}\vkl\X\vp, \lb{+-DDF} \]
for $m\in\Zn$, $1\le i\le d-2$. Here we have used the well-known
Fubini--Veneziano coordinate and momentum fields, respectively,
\[ X^\mu(z) &:=&
   q^\mu-ip^\mu\ln z+i\sum_{m\ne0}\frac1m\ga^\mu_mz^{-m},
   \lb{FuVe-coo} \\
   P^\mu(z) &:=& i\frac{d}{dz}X^\mu(z)
              = \sum_{m\in\Zn}\ga^\mu_mz^{-m-1}, \lb{FuVe-mom} \]
and employed the standard normal-ordering
$\xord\ldots$ for the transversal DDF operators: 
\[ \xord{\Ai{m}\Aj{n}}
         := \cases{ \Ai{m}\Aj{n} & for $m\le n$, \cr
                      \Aj{n}\Ai{m} & for $m>n$. } \lb{nopx-1} \]

Let us recall from \ct{GeKoNi96} that the shift of any polarization vector
$\vgx^i(\va)\equiv\vgx^i(\vgl)$ along the \vgd direction leaves the
associated DDF operator $\Ai{m}(\va)$ unchanged for $m\ne0$, because
the residue of a total derivative always vanishes. On the other hand,
the difference $\vgx^i(\vgL)-\vgx^i(\vgl)$ for any $\vgl\in\gO(\vgL)$
is, without loss of generality, always proportional to $\vgd$. Thus, for
$m\neq 0$, we are effectively dealing with a single set of DDF
operators $\Ai{m}(\vgL)$ for the whole module $L(\vgL)$; the zero mode
operators do differ for different $\va$, however. For definiteness, we
choose the polarization vectors to be $\vgx^i(\vgL)$ throughout.

The above operators obey the commutation relations
\[ [\Ai{m},\Aj{n}] &=& m\gd^{ij}\gd_{m+n,0}\vkl\X\vp, \lb{DDFcom-1} \\[.5ex]
   [\AL{m},\Ai{n}] &=& 0, \lb{DDFcom-2} \\[.5ex]
   [\AL{m},\AL{n}] &=& (m-n) \AL{m+n} +
       \frac{26-d}{12}m(m^2-1) \gd_{m+n,0}\vkl\X\vp. \lb{DDFcom-3} \]
They are \emph{physical}, i.e.,
\[  [L_m,\Ai{n}]=[L_m,A^\pm_n] =0
    \qquad\forall m,n\in\Zn, 1\le i\le d-2, \nn \]
and therefore map physical into physical states. Moreover, they 
constitute a spectrum-gener\-ating algebra for the string. In particular,
\[ \cP[\svl]=\span\left\{\AI{i_1}{-m_1}\cdots\AI{i_M}{-m_M}
                     \AI{-}{-n_1}\cdots\AI{-}{-n_N}\ket{\va}\,\Big|\,
                     \sum m_\mu+\sum n_\nu=1-\frc12\vgl^2\right\},
    \lb{phys-span} \]
for a DDF decomposition $\vgl=\va-n\vkl$ of \vgl and for
$i_\mu=1,\ldots,d-2, m_\mu>0, n_1\ge\ldots\ge n_N\ge2$. Note that
$\AI{-}{-1}\ket{\va}\propto L_{-1}\ket{\va-\vkl}$, i.e., $\AI{-}{-1}$
generates null physical states which must be discarded.

\section{Affine vertex operators at arbitrary level} \lb{sec:Aff}
We introduce a linear map $\piL:\ \fg\ \longrightarrow \End\,\cPL$
as follows:
\[    K &\longmapsto& \vgd\X\vp, \non
    d &\longmapsto& \vgL_0\X\vp, \non
    H^i_m &\longmapsto&
     \Res{z}\vgx^i(\vgL)\X\vP(z)\exp\left[im\vgd\X\vX(z)\right], \non
    \Er{m} &\longmapsto&
     \Res{z}\ord{\exp\left[i(\vr+m\vgd)\X\vX(z)\right]}c_\vr,
    \lb{pi-def} \]
with $\vr\in\bar\gD$ and thus $\vr+m\vgd \in\gD$. $c_\vr$ denotes a
cocycle factor satisfying
$c_\vr\re^{i\vs\.\vq}=\gep(\vr,\vs)\re^{i\vs\.\vq}c_\vr$
for some bimultiplicative 2-cocycle $\gep$ normalized s.t.\
$\gep(\vo,\vo)=\gep(\vr,-\vr)=1$.
Indeed, it is straightforward to check (see \ct{Fren85,GodOli85}) that
the above operators are physical, i.e.,
\[ [L_m,\piL(K)]=[L_m,\piL(d)]=[L_m,\piL(H^i_n)]=[L_m,\piL(\Er{n})]=0 \nn \]
for all $m,n\in\Zn$, $\vr\in\bD$, $1\le i\le d-2$. More precisely, for
any $\vgl\in\gO(\vgL)$ one has
\[ \piL(K):\ \cP[\svl] &\longrightarrow& \cP[\svl], \non
   \piL(d):\ \cP[\svl] &\longrightarrow& \cP[\svl], \non
   \piL(H^i_m):\ \cP[\svl] &\longrightarrow& \cP[\svl+m\svd], \non
   \piL(\Er{m}):\ \cP[\svl] &\longrightarrow& \cP[\svl+\vr+m\svd].
   \lb{pi-map} \]
Furthermore, the following relations hold:
\[ [\piL(H^i_m),\piL(H^j_n)]
   &=& \cl m\gd^{ij}\gd_{m+n,0} , \lb{CWcom-1} \\[.5ex]
   [\piL(H^i_m),\piL(\Er{n})]
   &=& (\vgx^i\X\vr)\piL(\Er{m+n}), \lb{CWcom-2} \\[.5ex]
   [\piL(\Er{m}),\piL(\Es{n})]
   &=& \cases{ 0 & if $\vr\X\vs\ge0$, \cr
              \gep(\vr,\vs)\piL(\Erb{m+n}) & if $\vr\X\vs=-1$, \cr
              \piL(H^{\vr}_{m+n})+\cl m\gd_{m+n,0}  & if
              $\vr\X\vs=-2$,} \lb{CWcom-3} \\[.5ex]
   [\piL(K),\piL(x)]
   &=& 0 \qquad\forall x\in\fg, \lb{CWcom-4} \\[.5ex]
   [\piL(d),\piL(H^i_m)]
   &=& m\piL(H^i_m), \lb{CWcom-5} \\[.5ex]
   [\piL(d),\piL(\Er{m})]
   &=& m\piL(\Er{m}). \lb{CWcom-6} \]
Hence $\piL$ defines a level-\cl vertex operator realization of
\fg on \cPL. By identifying the vacuum vector $v_{\svL}$ in $L(\vgL)$
with the tachyonic ground state $\ket{\vgL}$ in \cPL, we conclude that
\[ L(\vgL)\hookrightarrow\cPL. \]
To see this, we first write down the realization of the
Chevalley--Serre generators, viz.
\[ e_i&:=&\piL(E^{\vr_i}_0),\quad
   f_i:=-\piL(E^{-\vr_i}_0),\quad
   h_i:=\piL(\vr_i\X\mathbf{H}_0)=\vr_i\X\vp,\quad
        \mbox{for\ }1\le i\le d-2, \non
   e_0&:=&\piL(E^{-\svt}_1),\quad
   f_0:=-\piL(E^{\svt}_{-1}),\quad
   h_0:=\piL(K-\vgt\X\mathbf{H}_0)=\vr_0\X\vp, \nn \]
where $\vgt$ denotes the highest root in \bD.
Then we have to verify (see e.g.\ \cite[Coroll.10.4]{Kac90}) both the
vacuum vector conditions
\[ e_I\ket{\vgL}=0\quad\mbox{for\ }0\le I\le d-2, \lb{vac-cond} \]
and the null vector conditions
\[ f_I^{1+\vr_I\.\svL}\ket{\vgL}=0\quad\mbox{for\ }0\le I\le d-2.
    \lb{null-cond} \]
{}From Eq.\ \Eq{pi-map} we infer that
\[ e_I: \cP[\svl] \longrightarrow \cP[\svl+\vr_I],\quad
   f_I: \cP[\svl] \longrightarrow \cP[\svl-\vr_I],\quad
   \mbox{for\ }0\le I\le d-2. \nn \]
Hence $e_I\ket{\vgL}$ has at least eigenvalue 2 for $L_0$ because
$\vgL^2=\vr_I^2=2$ and $\vgL$ is dominant, so that
$(\vgL + \vr)^2\geq 4$; but this contradicts the
fact that $e_I\ket{\vgL}$ is a physical state (cf.\ Eq.\
\Eq{phys-def}), hence $e_I\ket{\vgL}$ must be zero. The null vectors
$f_I^{1+\vr_I\.\svL}\ket{\vgL}$ vanish by the same argument since
$\frc12[\vgL-(1+\vr_I\X\vgL)\vr_I]^2=2+\vgL\X\vr_I$. In other words,
the null vectors are really zero in our approach. 
The space $\cP[\svl]$ must not be confused with the weight
space $L(\vgL)_{\svl}$, the space of states with weight $\vgl$ in the
representation; rather, we have the (in general proper) inclusion
\[  L(\vgL)_{\svl} \hookrightarrow \cP[\svl].   \]
If we make use of the observation that in \Eq{pi-def}
only transversal linear combinations of the string oscillators and
consequently transversal DDF operators can occur, we conclude that
\[  \mult_{\svL}(\vgl)=\dim L(\vgL)_{\svl}
    \le \dim \cP[\svl]_{\mathrm{transv.}}=p_{d-2}(1-\frc12\vgl^2). \]
This is an universal estimate for the weight multiplicities of any
irreducible affine highest weight module which seems to be new.  For
$\cl=1$ this bound is known to be saturated \ct{FreKac80}; at higher
level, however, the formula may constitute only a crude upper bound.
In general, there are ``missing states'', namely the physical states
which lie in $\cP[\svl]$ but not in $L(\vgL)_{\svl}$. Note, however,
that these have nothing to do with the above null vectors.

We also note that
\[ \piL(H^i_m)=\Ai{\cl m}(\vgL), \lb{H=A} \]
which shows that the transversal DDF operators $\Ai{\cl m}(\vgL)$ occur not
only as part of the spectrum generating algebra for the physical
string states but also as homogeneous Heisenberg subalgebra of the
affine algebra.  One might therefore ask whether it is possible to
rewrite the step operators $\piL(\Er{m})$ also in a manifestly
physical form in terms of these DDF operators. Indeed, this
is possible if one in addition uses the Lorentz generators. To
this end we introduce the \textbf{tranversal coordinate field}
\[ \XI{\svL}(z)\equiv\XI{}(z):=
   (\vgx^i)_\mu(\vkl)_\nu M^{\mu\nu}-i(\vgx^i\X\vp)\ln z
   +i\sum_{m\ne0}\frac1m\Ai{m}(\vgL)z^{-m},
   \lb{DDF-coo} \]
where
\[ M^{\mu \nu}:= q^\mu p^\nu - q^\nu p^\mu - i\sum_{n\neq 0}
   \frac{1}{n} \alpha_{-n}^{[\mu} \alpha_n^{\nu ]} \lb{Mmunu} \]
are the Lorentz generators,
and the \textbf{transversal momentum field}
\[ \PI{\svL}(z)\equiv\PI{}(z)
            := i\frac{d}{dz}\XI{}(z)
             = \sum_{m\in\Zn}\Ai{m}(\vgL)z^{-m-1}, \lb{DDF-mom} \]
(where we again do not indicate the dependence on the level explicitly.)
Note that the center of mass coordinate in \Eq{DDF-coo} is 
$(\vgx^i)_\mu(\vkl)_\nu M^{\mu\nu}$ rather than $q^i$ as one might
have naively guessed. This choice is forced upon us by the
requirement that the field $\XI{}(z)$ should be physical: since
\[ [ L_m , M^{\mu \nu} ] =0, \]
we have, with the definition \Eq{DDF-coo},
\[ [L_m, \XI{}(z) ] =0, \lb{phys-coo}  \]
whereas \Eq{phys-coo} would not vanish if the zero mode were $q^i$. 
Secondly, substituting $(\vgx^i)_\mu(\vkl)_\nu M^{\mu\nu}$  for
$q^i$ amounts to a replacement of a translation generator (in
momentum space) by a Lorentz rotation.  As we will see, this is
precisely what is required because our new expressions are defined in
terms of DDF operators which shift the momentum by a vector
proportional to $\vkl$ whereas \Eq{pi-def} is defined in terms of
ordinary string oscillators $\ga_m^\mu$ which do not shift momentum.
There is a corresponding reinterpretation of the affine Weyl
translation by a Lorentz boost (see Sect.\ \ref{sec:Wey}).

Observe also that we have defined the new field $\XI{}(z)$ so far only
for transversal indices. However, our definition can be generalized to
the longitudinal components $\sX^\pm$ by means of the operators
$\AI\pm{m}$ defined in Eqs.\ \Eq{l-DDF} and \Eq{+-DDF}, respectively.
Note that $\AP{m}\equiv 0$ for all $m\neq 0$, just as in light cone
gauge string theory. Although we will make no use of the components
$\sX^\pm$ in this paper, we expect them to become relevant in future
generalizations involving level-changing operators (see Sect.\
\ref{sec:Lon}). We note that the fields \Eq{DDF-coo} are
transcendental expressions in terms of the standard oscillator basis.

Next we establish the relation between the ``old'' step operators
$\piL(\Er{m})$ defined in \Eq{pi-def} and a set of new ones which
manifestly depend on the DDF operators. The ``new'' level-\cl
step operators are defined by
\[ \Ehr{m}(\vgL) :=\Res[0]{z} z^{\cl m}
  \xord{\exp\left[i\vr\X\vcX(z)\right]}c_\vr, \lb{Ehat}   \]
where we use the standard normal ordering \Eq{nopx-1} for the
Heisenberg oscillators and where the cocycle factors $c_\vr$, which
are functions of momentum, were explained after \Eq{pi-def} and
are the same as in \ct{FreKac80}. The
operators \Eq{Ehat} will permit us to evaluate the action of the step
operators directly in terms of the DDF basis.

\begin{Thm} \lb{Thm1}
On the representation space $L(\vgL)$, we have
\[ \piL(\Er{m})=\Ehr{m}(\vgL), \]
where the operators $\piL(\Er{m})$ and $\Ehr{m}(\vgL)$ are defined,
respectively, in \Eq{pi-def} and \Eq{Ehat}.
Consequently, the operators $\Ehr{m}(\vgL)$ , $\Ai{\cl m}(\vgL)$,
$\vgd\X\vp$, and $\vgL_0\X\vp$ realize the affine algebra at
level \cl on $\cPL$ in terms of the transversal Heisenberg algebra
spanned by the $\Ai{m}$'s.
\end{Thm}

\noi{\it Proof:}
By construction, the operators $\Ehr{m}$ are physical. The DDF
operator \Ai{n} shifts the momentum by $n\vkl$, and since the residue
in \Eq{Ehat} picks up $1+\cl m+\vr\X\va$ of such modes for \Ehr{m}
($\va$ denotes the eigenvalue of $\vp$), the contribution of the DDF
oscillators to the shift of momentum will be $(1+\cl m+\vr\X\va)\vkl$.
On the other hand, the zero mode involving the Lorentz generators
provides a momentum shift by $\vr- \left(1+\vr\X\va\right)\vkl\equiv
\tr{\vr}(\va)- \va$, so that in total $\Ehr{m}$ maps $\cP[\svl]$ into
$\cP[\svl+\vr+m\svd]$ as required. The momentum shift $\tr{\vr}(\va)$
is just a Weyl translation, and we will return to this point in Sect.\
\ref{sec:Wey} (see Eq.\ \Eq{a'}).

Next, we have to check that the new step operators satisfy the
required commutation relations, and this part of the proof is
very similar to the corresponding one for the FKS construction.
{}From the last observation we immediately get
\[ [\vv\X\vp,\Ehr{m}]=\vv\X(\vr+m\vgd)\Ehr{m}, \nn \]
for any $\vv\in\fh^*$, which yields the correct commutation relations
with $\piL(K)$, $\piL(d)$, and $\piL(H^i_0)$. By the use of Eqs.\
\Eq{H=A}, \Eq{Weyl4} and \Eq{DDFcom-1} we obtain, for $m\ne0$,
\[ [\piL(H^i_m),i\XL{j}{\svL}(z)]=\gd^{ij}z^m, \nn \]
from which \Eq{CWcom-2} follows. Now, let us work out the commutator
of two step operators which amounts to calculating the operator
product of normal-ordered exponentials of the transversal coordinate
field, namely,
\[ [\Ehr{m},\Ehs{n}]=\gep(\vr,\vs)\Res[0]{w}\Res[w]{z}z^{\cl m}w^{\cl n}
                     \xord{\exp\left[i\vr\X\vcX(z)\right]}
                     \xord{\exp\left[i\vs\X\vcX(w)\right]}c_{\vr+\vs}. \nn \]
We split the transversal coordinate field as follows:
\[ \XI{\svL}(z)= \XI<(z)+Q^i-iP^i\ln z+\XI>(z), \nn \]
where
\[ \XI\lessgtr(z):=i\sum_{m\lessgtr0}\frac1m\Ai{m}(\vgL)z^{-m},\qquad
   Q^i:=(\vgx^i)_\mu(\vkl)_\nu M^{\mu\nu},\qquad
   P^i:=\vgx^i\X\vp, \nn \]
so that we can write the step operators explicitly as
\[ \Ehr{m}=\Res{z}\exp\left[i\vr\X\vcX_<(z)\right]
                  \re^{i\vr\.\vQ}z^{\vr\.\vP}
                  \exp\left[i\vr\X\vcX_>(z)\right]c_\vr. \nn \]
Using \Eq{DDFcom-1} and the relation $[Q^i,P^j]=i\gd^{ij}$, which is
valid on the level-\cl subspace $\cPL$ only, we find that
\[ \exp\left[i\vr\X\vcX_>(z)\right]
   \exp\left[i\vs\X\vcX_<(w)\right]
   &=& \left(1-\frac{w}{z}\right)^{\vr\.\vs}
   \exp\left[i\vs\X\vcX_<(w)\right]
   \exp\left[i\vr\X\vcX_>(z)\right]\quad(|z|>|w|), \non[.5ex]
   z^{\vr\.\vP}\re^{i\vs\.\vQ}
   &=&z^{\vr\.\vs}\re^{i\vs\.\vQ}z^{\vr\.\vP}, \non[.5ex]
   \re^{i\vr\.\vQ}c_\vr\re^{i\vs\.\vQ}c_\vs
   &=&\gep(\vr,\vs)\re^{i(\vr+\vs)\.\vQ}c_{\vr+\vs}. \nn \]
Thus we have
\[ [\Ehr{m},\Ehs{n}]&=&
   \gep(\vr,\vs)\Res[0]{w}\Res[w]{z}\bigg\{z^{\cl m}w^{\cl n}
   (z-w)^{\vr\.\vs}
   \exp\left[i\vr\X\vcX_<(z)+i\vs\X\vcX_<(w)\right]\times \non
   & &\hh{34mm}\times\re^{i(\vr+\vs)\.\vQ}
   z^{\vr\.\vP}w^{\vs\.\vP}
   \exp\left[i\vr\X\vcX_>(z)+i\vs\X\vcX_>(w)\right]\bigg\}c_{\vr+\vs}. \nn \]
It is clear that the commutator vanishes for $\vr\X\vs\ge0$. For
$\vr\X\vs=-1$, we have $\vr+\vs\in\bD$. Furthermore, the contour
integral of $z$ around $w$ then has the effect of setting $z=w$ in the
integrand due to the simple pole, and the result
$\gep(\vr,\vs)\Ehrs{m+n}$ follows. The case $\vr\X\vs=-2$ is equivalent
to $\vs=-\vr$ and corresponds to a second order pole at $z=w$ of the
integrand. Cauchy's theorem then yields the required result, viz.
\[ [\Ehr{m},\Ehs{n}]
   &=&\Res[0]{w}w^{\cl n}\dz
      \left[ z^{\cl m}
             +i\vr\X\vcX_<(z)
             +\left(\frac{z}{w}\right)^{\vr\.\vP}
             +i\vr\X\vcX_>(z)\right]_{z=w} \non
   &=&\Res[0]{w}w^{\cl(m+n)}
      \left[\vr\X\vcP(w)+\cl m w^{-1}\right] \non
   &=&\Ar{m+n}+\cl m \gd_{m+n,0}. \nn \]

Finally, we have to verify that \Ehr{m} really gives the same result
as $\piL(\Er{m})$ in terms of the string oscillators.  For this
purpose, we re-express the DDF operators for a given transversal
physical state in terms of ordinary string oscillators. Then the
leading oscillator contributions are the same because any product of
DDF operators $\Ai{m}$ differs from the corresponding product of
string oscillators $\ga_m^i$ only by terms all of which involve at
least one light-like oscillator $\vgd\X\vga_{-n}$ with the null root
$\vgd$.\footnote{This statement is no longer true for longitudinal DDF
  operators as can be seen by simple inspection of the examples
  (A.7)--(A.10) given in \ct{GebNic95}.}
Moreover, the Lorentz generators do not contribute to the
leading oscillator terms. Since any physical state is uniquely
determined by its leading oscillators, the result follows.
\begin{flushright} $\blacksquare$ \end{flushright}

We emphasize again that the equality stated in the theorem holds only on
the subspace $\cPL\subset{\cal P}$, but not on the whole physical 
state space $\cal P$. This is because we must utilize the relation
$\vkl\X\va=\vkl\X\vgl=1$ for all $\vgl\in\gO(\vgL)$ in the proof. We
note that an analogous result would hold for the original FKS
construction if one makes the simple replacement $z^m \rightarrow
z^{\cl m}$ (see \ct{Maro93} where this observation was made in the
context of parafermions). However, in our approach, the level-\cl
representation space is a subspace of a much bigger space, the space
of all physical states of arbitrary level, which enables us
to treat infinitely many irreducible representations
simultaneously (as required by a representation theoretic approach to
hyperbolic Kac--Moody algebras).

One of the nice features of the above realization is that it allows us
to simply understand what is special about the basic representation,
i.e., level $\cl =1$: only in this case is it possible to express the
step operators entirely in terms of the homogeneous Heisenberg
subalgebra spanned by the $\piL(H^i_m)$'s; for $|\cl|>1$, the step
operators cannot be expressed in terms of the operators $\Ai{\cl m}$
alone.  As a result we again have ``missing states'', namely physical
states which cannot be ``reached'' by applying step operators
successively to the tachyonic vacuum vector $v_{\svL}\equiv\ket{\vgL}$
that defines the representation. As we pointed out already, these
``missing states'' must not be confused with the null vectors of the
conventional approach. The consequences of our new formulation for the
computation of affine characters is an intriguing problem for further
study.

As an application of Thm.\ \ref{Thm1} we can immediately rederive the
new expression for the affine Sugawara generators given in
\ct{GeKoNi96}. Recall that in terms of the affine Cartan--Weyl basis
\Eq{pi-def} these are given by
\[ \sL{m} := \frac1{2(\cl+\hv)}\sum_{n\in\Zn}
            \bigg[\sum_{i=1}^{d-2}\circord{\piL(H^i_n)\piL(H^i_{m-n})}
            +\sum_{\vr\in\bD}\circord{\piL(\Er{n})\piL(\Ema{m-n})}\bigg],
            \lb{Sug-def} \]
where \hv denotes the dual Coxeter number of $\bfg$. The new
normal-ordering symbol $\circord{\ldots}$ refers to the mode indices
of the affine generators; for the operators $\piL(H^i_n)$ (but not
for the step operators!) it is the same as \Eq{nopx-1} by \Eq{H=A}.
The operators $\sL{m}$ are well known to generate a Virasoro
algebra (see e.g.\ \ct{GodOli86} and references therein)
\[ [\sL{m},\sL{n}]=(m-n)\sL{m+n}
                   +\frac{c(\cl)}{12}(m^3-m)\gd_{m+n,0}\piL(K),
   \lb{Sug-Vir} \]
with central charge
\[ c(\cl):=\frac{\cl\dim\bfg}{\cl+\hv}. \lb{Sug-cc} \]
They act as outer derivations on the affine Lie algebra according to
\[ [\sL{m},\Ai{\cl n}]=-n\Ai{\cl(m+n)},\qquad
   [\sL{m},\Ehr{n}]=-n\Ehr{m+n}. \lb{sem-prod} \]
In particular, we observe that $\sL{0}=-\piL(d)$.
By construction, the Sugawara generators are physical, viz.
\[ [\sL{m},L_n]=0\qquad\forall m,n\in\Zn. \]
Now let $\gz := e^{\frac{2\pi i}{\cl}}$ (or any other primitive
$\cl$-th root of unity). We have \ct{GeKoNi96}
\begin{Cor}
The operators $\sL{m}$ can be directly expressed in terms of
the transversal Heisenberg algebra by
\[ \sL{m}
        &=& \frac1{2\cl}\sum_{n\in\Zn}
              \sum_{i=1}^{d-2}\xord{\Ai{\cl n}\Ai{\cl(m-n)}}
            +\frac{\hv}{2\cl(\cl+\hv)}\sum_{n\neq 0(\cl)}
              \sum_{i=1}^{d-2}\xord{\Ai{n}\Ai{\cl m-n}}
             +\frac{(\cl^2-1)(d-2)\hv}{24\cl(\cl+\hv)}\gd_{m,0} \non
        & &{}-\frac1{2\cl(\cl+\hv)}\sum_{\vr\in\bD}
              \sum_{p=1}^{\cl-1}\frac1{|\gz^p-1|^2}
              \Res[0]{z}z^{\cl m -1}
              \xord{\exp\left\{i\vr\X
                               \left[\vcX(\gz^pz)-\vcX(z)\right]
                               \right\}}. \lb{Sug} \]
\end{Cor}

\noi{\it Proof:} Using the operator expansion in the proof of Thm.\
\ref{Thm1} we get
\[ \lefteqn{\sum_{n\in\Zn} \circord{\Ehr{-n}\Ehmr{m+n}} \equiv
   \sum_{n\ge0}\Ehr{-n}\Ehmr{m+n}+
            \sum_{n>0}\Ehr{m-n}\Ehmr{n}} \hh{8mm} \non
   &=&
    \bigg\{\Res{z}\Res[|z|>|w|]{w}-\Res{z}\Res[|z|<|w|]{w}\bigg\}
    (z-w)^{-2}\sum_{n\ge0}z^{-\cl n}w^{\cl(m+n)}
    \left(\frac{z}{w}\right)^{\vr\.\vP}\times \non[.5ex]
   & & \hspace{40mm} \times
    \exp\left\{i\vr\X\left[\vcX_<(z)-\vcX_<(w)\right]\right\}
    \exp\left\{i\vr\X\left[\vcX_>(z)-\vcX_>(w)\right]\right\} \non[1ex]
   &=&
    \Res[0]{w}\sum_{p=1}^{\cl}\Res[w_p]{z}\left\{
    \frac{z^\cl w^{\cl m}}{(z-w)^2(z^\cl-w^\cl)}
    \xord{\exp\left\{i\vr\X\left[\vcX(z)-\vcX(w)\right]\right\}}
    \right\}, \nn \]
where $w_p:=\gz^pw$. With the identity
\[ \frac{z^\cl-w^\cl}{z-w_p}=
   z^{\cl-1}+ z^{\cl-2}w_p+ \ldots +zw_p^{\cl-2}+ w_p^{\cl-1}
   =: F(z,w_p), \nn \]
the sum over the poles at $z=w_p$ for $1\le p\le\cl-1$
immediately yields the third term in \Eq{Sug}. As regards
the pole at $z=w$, we have to evaluate
\[ \lefteqn{\sum_{\vr\in\bD}\Res[0]{w}
   \frac12\frac{d^2}{dz^2}\left\{
    \frac{z^\cl w^{\cl m}}{F(z,w_\cl)}
    \xord{\exp\left\{i\vr\X\left[\vcX(z)-\vcX(w)\right]\right\}}
    \right\}\bigg|_{z=w}} \hh{8mm} \non
   &=&
   \frac12\sum_{\vr\in\bD}\Res[0]{w}\left\{
      w^{\cl m}\frac{d^2}{dz^2}
         \left[\frac{z^\cl}{F(z,w_\cl)}\right]_{z=w}
     +\frac1\cl w^{\cl m+1}\xord{
         \left[\frac{d}{dw}i\vr\vcX(w)\right]^2}\right\}, \nn \]
where the terms linear in $\vr$ drop out due to the sum over both positive
and negative roots. Using \Eq{DDF-mom} and the fact that
\[ \sum_{\vr\in\bD}\vr\XO\vr=2\hv\sum_{i=1}^{d-2}\vgx^i\XO\vgx^i,
   \nn \]
the second term is seen to give a contribution
\[ \frac{\hv}{2\cl(\cl+\hv)}\sum_{n\in\Zn}
              \sum_{i=1}^{d-2}\xord{\Ai{n}\Ai{\cl m-n}} \]
in the formula for the Sugawara operators. Finally, a straightforward
calculation yields
\[ \frac12\frac{d^2}{dz^2} \left[\frac{z^\cl}{F(z,w_\cl)}\right]_{z=w}
   =\frac{\cl^2-1}{12\cl w} \nn \]
which, together with $|\bD|=(d-2)\hv$, leads to the constant term in
\Eq{Sug}.
\begin{flushright} $\blacksquare$ \end{flushright}

\section{The affine Weyl group} \lb{sec:Wey}
We now return to the remarks made at the beginning of the proof
of Thm.\ \ref{Thm1}. As noted there, the momentum shift effected by the
step operator \Eq{Ehat} on a given state is the combination of
the shifts effected by the DDF operators (which are always along
the null root $\vgd$) and a zero mode contribution, such that the
total shift coincides with the one obtained for the original
step operator of \Eq{pi-map}. Furthermore, we observed that
the so-called (affine) Weyl translations naturally appeared there;
the latter are designated by $t_\vr\in\fT$ for $\vr\in\bar{Q}$
and act on $\fh^*$ as
\[ t_\vr(\vv)
   := \vv + (\vv\X\vgd)\vr -
      \left[(\vv\X\vgd)\frc12\vr^2 + \vr\X\vv\right]\vgd,
      \lb{Weyl0}  \]
where $\vv\in\fh^*$. Now, it is a well-known result that the affine
Weyl group is the semi-direct product of the Weyl group of the
underlying finite dimensional Lie algebra \bfg and the affine Weyl
translations, i.e.
\[ \fW(\fg)= \fW(\bfg)\ltimes\fT . \lb{Weyl-trans}  \]
To re-examine this result in the light of our approach we need the
following family of translations:
\[ \tr{\vr}(\vv) \equiv \vv'
   := \vv + (\vv\X\vkl)\vr -
      \left[(\vv\X\vkl)\frc12\vr^2 + \vr\X\vv\right]\vkl,
      \lb{Weyl1}  \]
for $\cl\in\Nn$ and $\vkl$ was defined in \Eq{k-def}. More specifically,
we have the following transformation formulas for a tachyonic
level-\cl vector $\va$, a polarization vector $\vgx^i(\va)$, and the
affine null vector \vgd, respectively:
\[ \tr{\vr}(\va) \equiv \va'
   &=& \va+\vr-\left(\frc12\vr^2+\vr\X\va\right)\vkl, \lb{a'} \\
   \tr{\vr}(\vgx^i(\va)) \equiv \vgx^i(\va')
   &=& \vgx^i(\va)-\left(\vr\X\vgx^i(\va)\right)\vkl, \lb{x'} \\
   \tr{\vr}(\vgd) \equiv \vgd'
   &=& \vgd. \lb{d'} \]
The above maps are linear and indeed fulfill the translation property
\[ \tr{\vr}\circ\tr{\vs}=\tr{\vr+\vs}\qquad\forall\vr,\vs\in\bar{Q}. \]
Moreover, they preserve the (Minkowskian!) norm, i.e. ${\vv'}^2 =
\vv^2$. We will now exploit this fact by re-interpreting them as
Lorentz boosts. In this way the affine Weyl group becomes a discrete
subgroup of ${\rm ISO} (d-2)$, the subgroup of the full Lorentz group
${\rm SO}(d-1,1)$ leaving fixed a given light-like vector. For the
level-preserving transformations considered in this section,
${\rm ISO}(d-2)$ is therefore nothing but the stability subgroup (in the
hyperbolic Weyl group) of the affine null root $\vgd$.

To proceed, we rewrite \Eq{Weyl1} as
\[ v_\mu' =  {\sT_\mu}^\nu v_\nu   \lb{Weyl2}   \]
with
\[ {\sT_\mu}^\nu \equiv {(\tr{\vr})_\mu}^\nu
  := \gd_\mu^\nu + r_\mu (\vkl)^\nu - (\vkl)_\mu r^\nu
  -\frc12 \vr^2 (\vkl)_\mu (\vkl)^\nu.   \lb{Weyl3}   \]
It is elementary to show that $\sT = \exp \gol$ with
\[ \gol_{\mu \nu}:= r_\mu (\vkl)_\nu - r_\nu (\vkl)_\mu.   \lb{omega}   \]
We have a unitary representation $\hsT$ of this Lorentz boost
on the Fock space by means of the Lorentz generators \Eq{Mmunu}, viz.
\[ \hsT:=\exp \left(\frac{i}2 \gol_{\mu\nu}M^{\mu\nu}\right); \lb{boost-rep} \]
one finds that
\[ \hsT(\vgl\X\vq)(\hsT)^{-1}=\tr{\vr}(\vgl)\X\vq,\qquad
   \hsT(\vgl\X\vga_m)(\hsT)^{-1}=\tr{\vr}(\vgl)\X\vga_m
   \quad\forall m\in\Zn\,. \lb{osc-trafo} \]
For instance, on the tachyon state $\ket\va$, we get
\[ \hsT \ket\va = \ket{\va'}. \]
Since, as we already noted, the transverse DDF oscillators
remain unchanged for $m\neq 0$, we therefore have, on $\cPL$,
\[ \hsT\Ai{m}(\va)(\hsT)^{-1}=\Ai{m}(\va')
    &=&\Ai{m}(\va), \lb{Weyl4} \\
   \hsT\Ai{0}(\va)(\hsT)^{-1}=\Ai{0}(\va')
    &=&\Ai{0}(\va)-\vr\X\vgx^i(\va). \lb{Weyl5}  \]

Although the replacement of the momentum shift by the Weyl
translation \Eq{Weyl1} and the re-interpetation of this translation
as a Lorentz boost was forced on us by the replacement
of ordinary string oscillators by DDF operators, it is now clear
that this interpretation is the natural one. This is also evident
from the following diagram, which displays the nested sequence
of Weyl groups of the finite, affine and hyperbolic Kac--Moody algebras
$\bfg \subset \fg \subset \bfgh$ as discrete subgroups of the
corresponding continuous groups:
\[ \begin{array}{ccccc}
  \fW (\bfg) &\subset &\fW (\fg) &\subset &\fW (\bfgh )      \\[1ex]
   \cap &  &\cap  & &  \cap  \\[1ex]
 {\rm SO}(d-2) &\subset &{\rm ISO}(d-2) &\subset &{\rm SO}(d-1,1)
   \end{array}  \lb{diagram}    \]
We can thus think of the affine Weyl group as a ``dimensional
null reduction" of the full hyperbolic Weyl group, similar
in spirit to the Kaluza--Klein reduction of Einstein's theory with a
null-Killing vector which was recently studied in \ct{JulNic95}, where the
group ${\rm ISO}(d-2)$ made its appearance as the residual tangent
space symmetry.

\section{Longitudinal DDF operators and level-changing
   tra\-nsformations} \lb{sec:Lon} 
We finally turn to longitudinal DDF operators. In the representation
theory of affine algebras these have played no role so far, because
one usually considers only one representation at a time. By contrast,
the longitudinal DDF operators do change the level, and therefore
interpolate between different, and inequivalent affine
representations.  This fact is immediately evident if one allows the
vector $\vr$ in the exponent of \Eq{Ehat} to have level $\cl \neq 0$,
in which case the field $\sX^\mu$ inevitably acquires a longitudinal
component.  Conversely, a non-vanishing longitudinal component in this
expression implies that $\vr$ cannot only have components in the
affine root lattice, but must have $\cl \neq 0$. While the necessity
of studying several representations simultaneously does not arise in
the theory of affine representations as such, the problem must be
faced when one considers hyperbolic Kac--Moody algebras which contain
infinitely many affine representations of arbitrary level.
Furthermore, as shown in \ct{GebNic95}, longitudinal states do
appear in these algebras. We claim that the formalism developed in
this paper furnishes the requisite tools for further investigations
in this direction, because it allows us to embed the different
representations into a single Fock space of physical states. In this
section we present some preliminary results concerning the
longitudinal DDF operators, which we expect to become relevant in
future developments. In particular, we will also consider the
level-changing generalizations of the Lorentz boosts \Eq{Weyl3}, whose
unitary implementation yields operators interpolating between DDF
operators of different level.

Because there is an infinity of tachyonic states $\ket\va$, the
longitudinal DDF operators $\AL{m}(\va)$ introduced in \Eq{l-DDF}
constitute an infinity of Virasoro algebras, but with uniform central
charge $c=26-d$, see \Eq{DDFcom-3}. By construction, all of these
commute with the Virasoro generators $L_m$, and are therefore
physical. Moreover, they also commute with the Sugawara generators
\Eq{Sug-def}
\[  [ \sL{m}, \AL{n}(\va) ] = 0      \]
for all tachyonic $\va$ associated with a weight $\vgl\in\gO(\vgL)$.
Although their polarization is along $\va$, a short calculation shows
that they are still invariant under the Lorentz boost \Eq{Weyl2}, in
accordance with Eqs.\ \Eq{Weyl4} and \Eq{Weyl5}; namely, we have
\[ \hsT\AL{m}(\va)(\hsT)^{-1}=\AL{m}(\va')
    &=&\AL{m}(\va), \non
   \hsT\AL{0}(\va)(\hsT)^{-1}=\AL{0}(\va')
    &=&\AL{0}(\va)+\vr\X\va. \nn  \]
Their commutation relations with the step operators \Eq{Ehat}
(for $\vr\in\bD$) are given by
\[ \Ehr{m} \AL{n}(\va)= \AL{n}(\va') \Ehr{m},    \]
where $\va'$ is the Weyl-boosted tachyon momentum defined in \Eq{a'}.
In deriving this result, the $A^iA^i$ term in \Eq{l-DDF} is essential.
The longitudinal DDF operators can thus be regarded as
intertwining operators between different (but isomorphic)
representations. The above relation also permits us to extend the
proof of Thm.~\ref{Thm1} to states containing longitudinal
excitations by simply moving all step operators to the right of the
longitudinal DDF operators.

So far, we have only considered the action of integrated tachyon
vertex operators associated with affine roots $\vr+m\vgd\in\gD$. From
the point of view of string theory it is natural to incorporate
``step operators'' associated with arbitrary tachyonic affine
dominant integral weights $\vgL'$. So let us define
\[ E^{\svL'}:=\Res{z}\ord{\exp\left[i\vgL'\X\vX(z)\right]}c_{\svL'} \]
for $\vgL'\in\rP_+$ satisfying $\vgL'^2=2$. Since $\vgL'$ has
non-vanishing level in general, only the special case of level zero
($\vgL'=\vr+m\vgd$ for $\vr\in\bD$) leads to the step operators
$\Er{m}$. By construction, the generalized step operators are
physical,
\[ E^{\svL'}: \cP[\svl] \longrightarrow \cP[\svl+\svL']; \]
but the crucial observation is that they change the level, i.e., they
map from a highest weight module $L(\vgL)$ to the module
$L(\vgL+\vgL')$ . Again, one might wonder whether these operators can
be rewritten in a manifestly physical form, and it is at this point
that the longitudinal DDF operators will enter the stage. In the
remainder, we will therefore generalize the affine Weyl translations
to level-changing translations which will be necessary for rewriting
the generalized step operators in terms of the DDF operators.

For any $\vgl_1$ in the weight system $\gO(\vgL_1)$ of a dominant weight
$\vgL_1$ with positive level $\cl_1$, we define
\[ \tr{\svl_1}(\vv)
   := \vv + (\vv\X\vkl)\vgl_1 -
      \left[(\vv\X\vkl)\frc12\vgl_1^2 + \vv\X\vgl_1\right]\vk_{\cl+\cl_1}.
      \lb{l-Weyl1}  \]
Note that the affine null root is no longer invariant but is
rescaled according to
\[ \tr{\svl_1}(\vkl)=\vk_{\cl+\cl_1}. \lb{kl-change} \]
The above maps are linear and again fulfil the translation property
\[ \tr[\cl+\cl_1]{\svl_2}\circ\tr{\svl_1}
   =\tr{\svl_1+\svl_2}. \]
Moreover, they preserve the norm, which allows us to rewrite them as
Lorentz boosts. To see this, we rewrite \Eq{l-Weyl1} as
\[ {[\tr{\svl_1}(\vv)]_\mu}^\nu v_\nu =  {\sT_\mu}^\nu v_\nu
   \lb{l-Weyl2}   \]
with
\[ {\sT_\mu}^\nu \equiv {(\tr{\svl_1})_\mu}^\nu
  := \gd_\mu^\nu + (\vgl_1)_\mu (\vkl)^\nu
                 - (\vk_{\cl+\cl_1})_\mu (\vgl_1)^\nu
                 - \frc12\vgl_1^2(\vk_{\cl+\cl_1})_\mu(\vkl)^\nu
   \lb{boost}. \lb{l-Weyl3} \]
A careful calculation shows that $\sT = \exp \gol$ with
\[ \gol_{\mu \nu}:= \ln\left(1+\frac{\cl_1}\cl\right)
   \left[(\vgl_1)_\mu (\vk_{\cl_1})_\nu
         - (\vk_{\cl_1})_\mu(\vgl_1)_\nu\right]. \lb{l-omega} \]
The last two equations generalize the expressions in \Eq{Weyl3} and
\Eq{omega}, respectively, which can be reobtained by putting
$\cl_1=0$. The unitary representation $\hsT$ of this level-changing
Lorentz boost on the Fock space is still given by formula
\Eq{boost-rep}. By Eq.~\Eq{kl-change}, conjugation with the operator
$\hsT$ transmutes level-\cl DDF operators into level-(\cl+$\cl_1$) DDF
operators.

A natural ansatz for a \textbf{longitudinal coordinate field}
analogous to \Eq{DDF-coo} is
\[ \XL-{\svL,\svL_1}(z):= \ln\left(1+\frac{\cl_1}\cl\right)
   (\vgL_1)_\mu (\vk_{\cl_1})_\nu M^{\mu\nu}
   -i(\vgL_1\X\vp)\ln z
   +i\sum_{m\ne0}\frac1m\AL{m}(\vgL+\vgL_1)z^{-m},
   \lb{l-DDF-coo} \]
with associated \textbf{longitudinal momentum field}
\[ \PL-{\svL,\svL_1}(z)
            := i\frac{d}{dz}\XL-{\svL,\svL_1}(z)
             = \sum_{m\in\Zn}\AL{m}(\vgL+\vgL_1)z^{-m-1}.
   \lb{l-DDF-mom} \]
This means, however, that one will have to face up to the problem of
dealing with exponentials of such operators. This is not quite the
same as exponentiating the Virasoro algebra because our operators are
always well-defined on finite occupation number states, for which the
contour integral picks up only finitely many contributions, but the
technical problems of manipulating such expressions (e.g.\ computing
operator products analogous to the ones used in the proof of
Thm.~\ref{Thm1}) still seem daunting. The relevant calculations would
be analogous to the computation of string scattering amplitudes in the
light cone gauge, with the longitudinal operators and bilinears of
transversal oscillators in the exponent --- something that apparently
has never been done in the literature.

\vspace*{4.5ex} \noi
\textbf{Acknowledgments:} We would like to thank J.~Fuchs for careful
reading of the manuscript.


\end{document}